# The elasto-/hydro-dynamics of quasicrystals with 12- and 18-fold symmetries in some soft matters and mathematical solutions

Tian You Fan

School of Physics, Beijing Institute of Technology, Beijing 100081, China

E-mail: tyfan@bit.edu.cn; tyfan2006@yahoo.com.cn

**Abstract**

The observation recently of 12-fold quasicrystals in polymers, nanoparticle mixture and 12-fold and 18-fold quasicrystals in colloidal solutions are important events for the study of quasicrystals. To describe the mechanical behaviour we propose a new solid-liquid phase quasicrystal model for some soft matters including polymers and colloids. The so-called new solid-liquid phase, is a new phase model of anisotropic fluid, but different from liquid crystal phase, here the structure presents quasiperiodic symmetry. Based on the model, the elasticity，fluidity and viscosity of the material have been studied, the relevant mathematical theory has also been proposed. Some mathematical solutions of the theory are discussed.

## 1. Introduction

The solid quasicrystals (in binary and ternary metallic alloys) including three-dimensional icosahedral quasicrystals [1] and two-dimensional 5-, 10-, 8- and 12-fold symmetry quasicrystals [2] were observed over twenty five years ago. Recently the structures were discovered in colloidal solutions [3], polymers [4-6] and nanoparticle mixture [7]. More recently the quasicrystals with 36-fold symmetry was created by nanofabricated technique [8]. Especially the work reported by Refs [3,8] are interesting in particular, because the 18-fold and 36-fold quasicrystals were first discovered. Before the experimental discovery of 12- and 18-fold quasicrystals, Denton and Loewen [9] theoretically predicted the existence of quasicrystals in colloids. From the view point of symmetry, the possible 7-, 14-, 9- and 18-fold quasicrystals in solid phase have been predicated by Hu et al [10] in an earlier time. However there is no further discussion on the structural and physical properties of the quasicrystals with 36-fold symmetry, this can be understood, that the observation was just reported.

The work reported in Refs [3-9] opens a fascinating area, concerning quasicrystals in soft matter. To develop the theory, we suggest a novel intermediate solid-liquid phase quasicrystals. This new theory may be a description of quasicrystals observed in some soft matters to date, advances widely the theoretical understanding of quasicrystals, and may present importance to engineering application of quasicrystals as a novel functional and structural material.

From the application angle, the mechanical behaviour of the quasicrystals with 12- and 18-fold symmetries in soft matter is one of basic problems. By using the so-called anisotropic fluid model, the elasticity, fluidity and viscosity of these two kinds of quasicrystals may be studied. It is fortunate the theoretical framework on the continuum mechanics of the new quasicrystalline phase can be drawn from the elasto-/hydro-dynamics of Lubensky et al [11-15], though the theory of Lubensky et al is directed to the quasicrystals in solid phase. However for 18-fold symmetry we must combine the theory of Lubensky et al with the use of the so-called "six-dimensional embedded space" concept proposed by Hu et al [10]. This combination allows us to set up a new theory in studying the novel quasicrystalline phase, at least for its macroscopic continuum mechanics. Physically and mathematically the problem is tremendous complex. Although there is a great difficulty, the static and some stationary dynamic problems can be solved. Some formulations on the elasticity and hydrodynamics of possible quasicrystals with 7-, 14-, 9- and 18-fold symmetries in solid phase have been discussed by Fan [16], one of purposes of whose work is attempted to study the relevant problems of quasicrystals in soft matter observed to date. According to supposition of Ref [16], the solutions for quasicrystals in solid phase e.g. collected in monograph [17] may be taken as the zero-order approximate solutions for the soft matter.

Considering the fact that the 12-fold quasicrystals in different soft matters were discovered frequently and the 18-fold quasicrystal in colloids first observed, we focus on the discussion on these two kinds of quasicrystals and their mechanical behaviour. As examples of solution of our theory, static and stationary dynamic problems in the quasicrystals are discussed.

## 2. New phase of anisotropic fluids for quasicrystals in some soft matters

In Ref [16] Fan discussed the elasticity and hydrodynamics of possible quasicrystals with 7-,14-,9- and 18-fold symmetries in solid phase. The work provides a hint for the study on mechanical behaviuor of quasicrystals in some soft matters.

Polymers, colloids and nanoparticle mixtures, in which the quasicrystals were observed, belong to very complicated phases due to the complex intra-structure of the materials. Drawing the experience in study of liquid crystals, we assume that some soft matters can be seen as a kind of anisotropic fluid macroscopically in which both elasticity and fluidity are in the same importance. This means the simplified phase has features with both fluids and solids. For simplicity, the fluid can be seen as a conventional incompressible viscous fluid, and the solid is an elastic body with certain quasiperiodicity. With this material model, the mechanics of 12-fold and 18-fold quasicrystals in some soft matters can be studied. Perhaps this is a theory of a lower order approximation of quasicrystals in real soft matter. The advantage of the theory lies in the quantitative stress analysis of elastic deformation and viscous flow is available.

**3. Elasticity, fluidity and viscosity of 12-fold quasicrystals in some soft matters**

Among the physical properties of quasicrystals in soft matters the elasticity, fluidity and viscosity are the most interested in by researchers. The simplified model of anisotropic fluid mentioned above is focused on these properties. The elasto-/hydro-dynamics originated from Lubensky et al [11] for quasicrystals in solid phase can be used for the present study as one of frameworks. We here give some modifications (or simplifications).The basic equations can be written as follows

$$\left. \begin{array}{l} \rho \dfrac{\partial V_i}{\partial t} + \rho V(\nabla V_i) = \dfrac{\partial}{\partial x_j}\left(\sigma_{ij} + \sigma'_{ij}\right) \\ \dfrac{\partial V_j}{\partial x_j} = 0 \\ \dfrac{\partial u_i}{\partial t} + V(\nabla u_i) + \Gamma_u \dfrac{\partial \sigma_{ij}}{\partial x_j} - V_i = 0 \\ \dfrac{\partial w_i}{\partial t} + V(\nabla w_i) + \Gamma_w \dfrac{\partial H_{ij}}{\partial x_j} = 0 \end{array} \right\} \quad (3\text{-}1)$$

in the equations $\rho$ denotes the mass density, $V=(V_x,V_y,V_z)$ the velocity vector, $\nabla$ the gradient operator, $\sigma_{ij}$ the stress tensor due to elastic deformation, $\sigma'_{ij}$ the fluid stress tensor, $u_i$ the phonon displacements(i.e., $u=(u_x,u_y,u_z)$ ), $w_i$ the phason displacements (i.e., $w=(w_x,w_y,w_z)$ ), $H_{ij}$ the phason stress tensor, $\Gamma_u$ the kinetic coefficient and $\Gamma_w$ the phason one, respectively. The elasto-/hydro-dynamic equations (3-1) present some differences from the original form given by Lubensky et al for describing quasicrystals of solid phase : 1) Physically the stress tensor $\sigma_{ij}'$ here, refer to equation (3-15), represents fluid stress tensor, while which in solid quasicrystals represents the stress tensor, refer to equation (10) of Ref [16], due to the solid viscosity ; 2) The computation shows the variation of the mass density $\rho$ is very small, we have assumed the medium is incompressible, so the mass density $\rho$ is a constant and does not consider it as a filed variable; 3) The hydrodynamic variable—the fluid pressure $p = p(x, y, z, t)$ is taken as an independent field variable of the medium.

According to the theory of elasticity of quasicrystals there are two strain tensors corresponding to phonon and phason fields, respectively

$$\varepsilon_{ij} = \frac{1}{2}\left(\frac{\partial u_i}{\partial x_j} + \frac{\partial u_j}{\partial x_i}\right), \quad w_{ij} = \frac{\partial w_i}{\partial x_j} \qquad (3\text{-}2)$$

The constitutive equations are

$$\left.\begin{aligned}\sigma_{ij} &= \frac{\partial F}{\partial \varepsilon_{ij}} = C_{ijkl}\varepsilon_{kl} + R_{ijkl}w_{kl} \\ H_{ij} &= \frac{\partial F}{\partial w_{ij}} = K_{ijkl}w_{kl} + R_{klij}\varepsilon_{kl}\end{aligned}\right\} \qquad (3\text{-}3)$$

where function $F = F(u, w) = F(\varepsilon_{ij}, w_{ij})$ is the elastic free energy of the system, $C_{ijkl}$ the elastic constant tensor of phonon field, $K_{ijkl}$ the one for phason field, $R_{ijkl}$ the one for phonon-phason coupling ($uw$ coupling) field. Here for 12-fold two-dimensional quasicrystals

$$R_{ijkl} = 0 \qquad (3\text{-}4)$$

because phonons and phasons are decoupled in the case.

For two-dimensional quasicrystals there are phonon components

$$u_x = u_x(x, y, z, t), u_y = u_y(x, y, z, t), u_z = u_z(x, y, z, t) \qquad (3\text{-}5)$$

and phason components

$$w_x = w_x(x, y, z, t), w_y = w_y(x, y, z, t), w_z = 0 \qquad (3\text{-}6)$$

where we assumed that the $z$ axis is the 12-fold symmetry axis of the material.

For simplicity we discuss only the plane fields in the following, i.e., in the case

$$\frac{\partial}{\partial z} = 0 \qquad (3\text{-}7)$$

For the quasicrystal of point group 12mm the elastic free energy in the plane elasticity

$$F(u, w) = F(\varepsilon_{ij}, w_{ij}) = \frac{1}{2} L(\nabla \cdot u)^2 + M \varepsilon_{ij} \varepsilon_{ij} + \frac{1}{2} K_1 w_{ij} w_{ij} + \frac{1}{2} K_2 (w_{21}^2 + w_{12}^2 + 2 w_{11} w_{22}) + \frac{1}{2} K_3 (w_{21} + w_{12})^2 = F_u + F_w \quad (x = x_1, y = x_2, i = 1, 2, j = 1, 2) \qquad (3\text{-}8)$$

where $F_u$ is the elastic free energy contributed by the phonon field, and $F_w$ one comes from the contribution of the phason field, and

$$C_{ijkl} = L\delta_{ij}\delta_{kl} + M(\delta_{jk}\delta_{jl} + \delta_{il}\delta_{jk}) \qquad (i, j, k, l = 1, 2) \qquad (3\text{-}9)$$

$$L = C_{12}, M = (C_{11} - C_{12})/2 = C_{66} \qquad (3\text{-}10)$$

$$\left.\begin{array}{l} K_{1111} = K_{2222} = K_1, K_{1122} = K_{2211} = K_2 \\ K_{1221} = K_{2112} = K_3, K_{2121} = K_{1212} = K_1 + K_2 + K_3 \end{array}\right\} \qquad (3\text{-}11)$$

and others are zero. The results can also be expressed by

$$K_{ijkl} = (K_1 - K_2 - K_3)(\delta_{ik} - \delta_{il}) + K_2\delta_{ij}\delta_{kl} +$$
$$K_3\delta_{il}\delta_{jk} + 2(K_2 + K_3)(\delta_{i1}\delta_{j2}\delta_{k1}\delta_{l2} + \delta_{i2}\delta_{j1}\delta_{k2}\delta_{l1}) \quad (i,j,k,l = 1,2) \quad (3\text{-}12)$$

For soft matters, the body presents fluidity and viscosity. The fluidity can be expressed by velocity vector for two-dimensional assumption

$$V_x = V_x(x,y,t), V_y = V_y(x,y,t), V_z = 0 \quad (3\text{-}13)$$

Consider the linear Newton fluid, there is strain rate tensor (or the deformation velocity tensor)

$$\dot{\xi}_{ij} = \frac{1}{2}\left(\frac{\partial V_i}{\partial x_j} + \frac{\partial V_j}{\partial x_i}\right) \quad (3\text{-}14)$$

and the constitutive law for incompressible fluid

$$\sigma'_{ij} = -p\delta_{ij} + 2\eta\dot{\xi}_{ij} \quad (3\text{-}15)$$

in which $p$ denotes the pressure of the soft matter as mentioned previously, $\eta$ the viscosity coefficient of the matter, respectively. The total phonon stress tensor is

$$(\sigma_{ij})_{total} = \sigma_{ij} + \sigma'_{ij} \quad (3\text{-}16)$$

in which $\sigma_{ij}$ is defined by (3-3) and $\sigma'_{ij}$ by (3-15), respectively.

4. Elasticity, fluidity and **viscosity of 18-fold quasicrystals in some soft matters**

Although the 18-fold quasicrystals belong to two-dimensional quasicrystals like 12-fold quasicrystals, they are quite different each other due to the symmetry. This leads to different structural and physical properties between them. For 18-fold quasicrystals there are two different phason fields apart from phonon field, and they all are two-dimensional. The detail about this can be found in Ref [16]. So that the field variables and field equations of quasicrystals with 18-fold symmetry are different from those of in the quasicrystals with 12-fold symmetry, i.e.,

$$\left.\begin{aligned}
&\rho\frac{\partial V_i}{\partial t}+\rho V(\nabla V_i)=\frac{\partial}{\partial x_j}\left(\sigma_{ij}+\sigma'_{ij}\right)\\
&\frac{\partial V_j}{\partial x_j}=0\\
&\frac{\partial u_i}{\partial t}+V(\nabla u_i)+\Gamma_u\frac{\partial \sigma_{ij}}{\partial x_j}-V_i=0\\
&\frac{\partial v_i}{\partial t}+V(\nabla v_i)+\Gamma_v\frac{\partial \tau_{ij}}{\partial x_j}=0\\
&\frac{\partial w_i}{\partial t}+V(\nabla w_i)+\Gamma_w\frac{\partial H_{ij}}{\partial x_j}=0
\end{aligned}\right\} \quad (4\text{-}1)$$

in which $\rho$, $V=(V_x,V_y)$, $\nabla=\left(\mathrm{i}\frac{\partial}{\partial x}+\mathrm{j}\frac{\partial}{\partial y}\right)$, $u=(u_x,u_y)$, $w=(w_x,w_y)$ and $\sigma'_{ij}$ are similar to those given in Section 3, but there is a new phason field $v=(v_x,v_y)$, we call it as the first phason field, and call $w=(w_x,w_y)$ as the second phason field, this leads to the associated strain tensor with the first phason field is

$$v_{ij}=\frac{\partial v_i}{\partial x_j} \quad (4\text{-}2)$$

the corresponding stress tensor is denoted by $\tau_{ij}$. The new elastic constitutive law is

$$\sigma_{ij} = \frac{\partial F}{\partial \varepsilon_{ij}} = C_{ijkl}\varepsilon_{kl} + r_{ijkl}v_{kl} + R_{ijkl}w_{kl}$$

$$\tau_{ij} = \frac{\partial F}{\partial v_{ij}} = T_{ijkl}v_{kl} + r_{klij}\varepsilon_{kl} + G_{ijkl}w_{kl} \quad (4\text{-}3)$$

$$H_{ij} = \frac{\partial F}{\partial w_{ij}} = K_{ijkl}w_{kl} + R_{klij}\varepsilon_{kl} + G_{klij}v_{kl}$$

where $F = F(u,v,w) = F(\varepsilon_{ij}, v_{ij}, w_{ij})$ the strain energy density

$$F(u,v,w) = F(\varepsilon_{ij}, v_{ij}, w_{ij}) = \frac{1}{2}L(\nabla \cdot u)^2 + M\varepsilon_{ij}\varepsilon_{ij} + T_1\left[(v_{11}+v_{22})^2 + (v_{21}-v_{12})^2\right]$$
$$+ T_2\left[(v_{11}-v_{22})^2 + (v_{21}+v_{12})^2\right] + K_1\left[(w_{11}+w_{22})^2 + (w_{21}-w_{12})^2\right] + T_2\left[(w_{11}-w_{22})^2 + (w_{21}+w_{12})^2\right] +$$
$$G[(v_{11}-v_{22})(w_{11}-w_{22}) + (v_{21}+v_{12})(w_{21}+w_{12})] = F_u + F_v + F_w + F_{vw} \quad (x = x_1, y = x_2, i = 1,2, j = 1,2)$$

(4-4)

, and elastic constant $C_{ijkl}, K_{ijkl}, R_{ijkl}$ defined as in previous section, and $T_{ijkl}$ the elastic constant tensors associated with the first phason tensor $v_{ij}$, and $r_{ijkl}$ the $uv$ coupling elastic constant tensor, $G_{ijkl}$ the $vw$ coupling elastic constant tensor, respectively. In (4-1) there is a new kinetic coefficient $\Gamma_v$ due to the phason field $v = (v_x, v_y)$

The other elastic constants defined by (3-9)-(3-12) hold for the 18-fold quasicrystals as well.

The phason elastic constant tensor $T_{ijkl}$ has non-zero independent components [10,16]

$$T_{1111} = T_{2222} = T_{2121} = T_1$$
$$T_{1122} = T_{2211} = -T_{2112} = -T_{1221} = T_2 \quad (4\text{-}5)$$

and other $T_{ijkl} = 0$, and the expression of them by tensor of four rank is

$$T_{ijkl} = T_1\delta_{ik}\delta_{jl} + T_2(\delta_{ij}\delta_{kl} - \delta_{il}\delta_{jk}) \quad (i,j,k,l = 1,2) \quad (4\text{-}6)$$

In addition the phonon and first phason are decoupled, this leads to

$$r_{ijkl} = 0 \quad (4\text{-}7)$$

and the phonon field and second phason field are decoupled too i.e.,

$$R_{ijkl} = 0 \quad (4\text{-}8)$$

and phason-phason coupling elastic constants are as follows

$$G_{ijkl} = G(\delta_{i1} - \delta_{i2})(\delta_{ij}\delta_{kl} - \delta_{il}\delta_{jk} + \delta_{iljk}) \quad (i,j,k,l=1,2) \quad (4\text{-}9)$$

these are originated from to Hu et al [10].

5. Solution example

Solutions for the above theory are very important from the fundamental points and applications. It is evident the analytic solutions for the problems are quite difficult to construct, the solving in general is taken numerical methods. However there are possibility to obtain analytic solutions for some simple cases. In this section we list some general solutions in integral expressions.

5.1 Static solution for 12-fold quasicrystals in soft matter

Here we discuss some solutions for 12-fold quasicrystals in soft matter only.

Since the velocities are very small, the nonlinear terms in (3-1) can be omitted, then the equations have been linearized. In addition, we consider the static state (i.e., terms concerning time derivative $\partial/\partial t$ are taken to be zero), then the equations can be simplified further. Substituting the deformation geometry equations (3-2) into constitutive relations (3-3), and (3-14) into (3-15), respectively, then into the equations of motion (3-1), after some derivations, we find that, the fluid motion equations are homogeneous, i.e., the equations governing fluid variables $V_x, V_y, p$ are independent from the elastic fields, they can be solved directly. The general solution for fluid field is

$$\left. \begin{array}{l} V_y(x,y) = \dfrac{1}{2\pi}\int_{-\infty}^{\infty}\overline{V}_y(\xi,y)\exp(-i\xi x)d\xi \\[6pt] V_x(x,y) = \dfrac{1}{2\pi}\int_{-\infty}^{\infty}\overline{V}_x(\xi,y)\exp(-i\xi x)d\xi \\[6pt] p(x,y) = \dfrac{1}{2\pi}\int_{-\infty}^{\infty}\overline{p}(\xi,y)\exp(-i\xi x)d\xi \\[6pt] \overline{V}_y(\xi,y) = A_1(\xi)\exp\left(-\sqrt{\xi^2 + \dfrac{1}{\Gamma_u \eta}}\,y\right) + A_2(\xi)\exp\left(\sqrt{\xi^2 + \dfrac{1}{\Gamma_u \eta}}\,y\right) \\[6pt] + A_3(\xi)\exp(-|\xi|)y + A_4(\xi)\exp(|\xi|)y \\[6pt] \overline{V}_x(\xi,y) = -\dfrac{i}{\xi}\overline{V}_y{'}(\xi,y) \\[6pt] \overline{p}(\xi,y) = -\dfrac{i}{\xi}\left[i\dfrac{\eta}{\xi}\left(\dfrac{d^2}{dy^2} - \xi^2\right) - i\dfrac{1}{\Gamma_u \xi}\right]\overline{V}_y{'}(\xi,y) \end{array} \right\} \quad (5\text{-}1)$$

in which $A_1(\xi), A_2(\xi), A_3(\xi), A_4(\xi)$ are unknown to be determined by boundary conditions. The solution explores profoundly the effect of dissipation parameters $\Gamma_u$ and $\eta$, they appear in the basic solution in the form of product $\Gamma_u \eta$.

The phonon elastic deformation equations derived from equations of motion (3-1) are non-homogeneous, i.e., the displacement field $u_i$ is dependent upon the flow field, one can solve them only by alternative method. The zero-order alternative solution corresponds to the homogeneous equations is

$$\left. \begin{array}{l} u_x^{(0)}(x,y) = -(L+M)\dfrac{1}{2\pi}\int_{-\infty}^{\infty}\xi^2\left\{-|\xi|\left[B_1(\xi)+B_2(\xi)y\right]\exp(-|\xi|y) + |\xi|\left[B_3(\xi)+B_4(\xi)y\right]\exp(|\xi|y)\right\}\exp(-i\xi x)d\xi \\[6pt] u_y^{(0)}(x,y) = -(L+3M)\dfrac{1}{2\pi}\int_{-\infty}^{\infty}\left\{\xi^2\left[B_1(\xi)+B_2(\xi)y\right]\exp(-|\xi|y) + \left[B_3(\xi)+B_4(\xi)y\right]\exp(|\xi|y)\right\}\exp(-i\xi x)d\xi \end{array} \right\}$$

(5-2)

where $B_1(\xi), B_2(\xi), B_3(\xi), B_4(\xi)$ are unknown to be determined by boundary conditions. The zero-order alternative solution does not contain effect of fluid field, and the first-order alternative solution is

$$u_x^{(1)}(x, y) = u_x^{(0)}(x, y) + u_x^*(x, y)$$
$$u_y^{(1)}(x, y) = u_y^{(0)}(x, y) + u_y^*(x, y)$$

(5-3)

in which the second terms in (5-3) corresponding to a special solution of non-homogeneous equation that related to the velocity components, so the total solution contains the effect of fluid field.

The phason elastic deformation equations derived from equations of motion (3-1) are homogeneous, i.e., the displacement field $w_i$ is independent from the phonon displacement field as well as the fluid field, one can solve them directly, and the solution is

$$w_x(x, y) = -(K_1 + K_3)\frac{1}{2\pi}\int_{-\infty}^{\infty} \xi^2 \left\{ -|\xi|[C_1(\xi) + C_2(\xi) y]\exp(-|\xi| y) + |\xi|[C_3(\xi) + C_4(\xi) y]\exp(|\xi| y) \right\} \exp(-i\xi x) d\xi$$

$$w_y(x, y) = -(2K_1 + K_2 + K_3)\frac{1}{2\pi}\int_{-\infty}^{\infty} \left\{ \xi^2 [C_1(\xi) + C_2(\xi) y]\exp(-|\xi| y) + [C_3(\xi) + C_4(\xi) y]\exp(|\xi| y) \right\} \exp(-i\xi x) d\xi$$

(5-4)

where $C_1(\xi), C_2(\xi), C_3(\xi), C_4(\xi)$ are unknown to be determined by boundary conditions.

The above procedure holds for 18-fold quasicrystals too, the fluid field solution and phonon displacement field solution are the same as those for 12-fold quasicrystals, the only difference lies in the phason displacement fields, where there are two phason fields, but are decoupled to phonon filed as well as fluid field.

5.2 Stationary dynamic solution

We consider a stationary dynamic problem, e.g. if there is a disturb source propagating with uniform speed $V_0$ in the medium, taking the Galilean transformation

$$x = x_1 - V_0 t, \quad y = y_1 \qquad (5\text{-}5)$$

in which $(x, y)$ represents a moving coordinate system, $(x_1, y_1)$ the fixed one, respectively, and solve the problem in the moving coordinate system. The Fourier analysis is still effective, and we find that the argument of exponential functions the basic solution is a complex number $\left(\xi^2 + \dfrac{1}{\Gamma_u \eta} + i \dfrac{\rho V_0}{\eta} \xi\right)$ rather than the real number $\left(\xi^2 + \dfrac{1}{\Gamma_u \eta}\right)$, in which the imaginary part contains factor $\dfrac{\rho V_0}{\eta}$. The imaginary part describes the stationary dynamic effect and the dissipation effect commonly, in which $\rho V_0$ can be understood the momentum of the disturb source. So that, the solving presents more complex feature than that in the Subsection 5.1.

## 6. Conclusion and discussion

The physical framework on elasticity and viscosity of 12-fold quasicrystals in some soft matters is taken from the theory of Lubensky et al [11] created for quasicrystals of solid phase, but we here make some modifications, this leads to analytic solution be available. The physical framework of elasticity and viscosity of 18-fold quasicrystals in some soft matters partly is originated from the theory of Lubensky et al [11] too, and partly is originated from the concept of "the six-dimensional embedding space" of Hu et al [10] since the symmetry in 18-fold quasicrystals is different from that of 5-, 8- , 10- and 12-fold quasicrystals. In the discussion we simplify the theory of Lubensky et al aimed to construct analytic solution. The solutions for velocity field, pressure field and phason displacement field are exact in statics, and but it is needed to do alternative computation for the solutions for phonon displacement field, the general zero order alternative solution is also constructed. For some complicated boundary value problems the solutions for phonon and phason displacement fields can refer to those data accumulated in quasicrystals of solid phase so far. The higher order alternative solutions for phonon field are difficult, which can be treated only by numerical method, in general. The solutions for dynamic problems are also available, but the computation is more complex.

**Acknowledgment**   This work is supported by the National Natural Science Foundation of China through grant 11272053. The author thanks Professor Zhang Ping Wen in Peking University and Professor Hu Chen Zheng in Wuhan University for help discussion.